# K-Nearest Neighbor Classification over Semantically Secure Encrypted Relational Data

Gunjan Mishra[1], Kalyani Pathak[1], Yash Mishra[2], Pragati Jadhav[1], Vaishali Keshervani[1]

[1] K.K.Wagh Institute of Engineering and Research (Savitribai Phule Pune University), Nasik - 422003, India

[2] JK Lakshmipat University, Jaipur - 302026, India

*Corresponding author email: mgunjan.18@gmail.com

**Abstract**

Data mining has various real-time applications in fields such as finance, telecommunications, biology, and government. Classification is a primary task in data mining. With the rise of cloud computing, users can outsource and access their data from anywhere, offloading data and it's processing to the cloud. However, in public cloud environments while data is often encrypted, the cloud service provider typically controls the encryption keys, meaning they can potentially access the data at any time. This situation makes traditional privacy-preserving classification systems inadequate. The recommended protocol ensures data privacy, protects user queries, and conceals access patterns. Given that encrypted data on the cloud cannot be directly mined, we focus on a secure k-nearest neighbor classification algorithm for encrypted, outsourced data. This approach maintains the privacy of user queries and data access patterns while allowing effective data mining operations to be conducted securely in the cloud. With cloud computing, particularly in public cloud environments, the encryption of data necessitates advanced methods like secure k-nearest neighbor algorithms to ensure privacy and functionality in data mining. This innovation protects sensitive information and user privacy, addressing the challenges posed by traditional systems where cloud providers control encryption keys.

## 1. Introduction

Cloud computing enables entities to outsource both their databases and data processing functionalities. The cloud offers mechanisms for querying and managing hosted databases. Outsourcing data allows owners to benefit from reduced data management costs, storage overhead, and improved services. However, the cloud cannot be fully trusted, making data confidentiality and query privacy challenging to maintain due to various documented security risks. For instance, in the event of a breach, any unencrypted sensitive data can be easily exposed to attackers. One effective way to protect the confidentiality of outsourced data is for data owners to encrypt it before outsourcing. This encryption ensures data confidentiality even if the cloud is compromised by external threats like hacking. However, a potentially curious or malicious cloud operator can track user queries and infer what the user is looking for, compromising query privacy. To preserve query privacy, authorized users must encrypt their queries before sending them to the cloud for evaluation. Additionally, the cloud can derive sensitive information from observing data access patterns during query processing, even if the data and queries are encrypted. Data access patterns refer to relationships between encrypted data observed by the cloud during query processing, such as which records are retrieved. It is crucial to keep both the data and users' input queries private during query processing. The key question is, "How can the cloud execute queries over encrypted data without decrypting them or compromising user privacy?" This question has led to a specific research area known as query processing over encrypted data.

## 2. Proposed System

We propose a novel privacy-preserving k-NN classification protocol, denoted as PPkNN. In this scenario, we assume that Sonia's database consists of n records, denoted by $T=\{t_1, t_2, \ldots, t_n\}$, each with m attributes. Here, $t_{i,j}$ represents the j-th attribute value of record $t_i$. Initially, Sonia encrypts her database attribute-wise, computing $E_{pk}(t_{i,j})$ for $1 \leq i \leq n$ and $1 \leq j \leq m$. The encrypted database is denoted by $E_{pk}(T)$. Sonia then outsources $E_{pk}(T)$ and the future query processing service to the cloud. We will assume that all attribute values and their Euclidean distances lie within $[0, 2^l)$, where *l* is a predefined integer. The goal of PPkNN is to efficiently and securely retrieve the top k records that are closest to the user's query.

To achieve this, we employ three additional protocols alongside PPkNN:
1. *Secure Squared Euclidean Distance (SSED) Protocol:* This protocol calculates the squared Euclidean distance between encrypted data points, ensuring that the computations remain secure and private, thus preventing any leakage of information during the distance computation process.

2. *Secure Bit-Decomposition (SBD) Protocol*: This protocol securely decomposes encrypted values into their constituent bits, facilitating further encrypted computations. This is crucial for performing operations that require bit-level manipulation without revealing the actual values.

3. *Secure Minimum out of n Numbers (SMINn) Protocol*: This protocol securely determines the smallest value among n encrypted numbers, enabling the identification of the nearest neighbors in the encrypted domain. This ensures that the closest records to the query are accurately and securely identified.

By utilizing these protocols, PPkNN ensures that Sonia's database and user queries remain confidential, while still allowing efficient and accurate k-NN classification on the encrypted data outsourced to the cloud. This approach mitigates the risks associated with outsourcing data to potentially untrusted cloud service providers by maintaining data confidentiality and query privacy through advanced cryptographic techniques. This proposed system is particularly beneficial for scenarios where data privacy is paramount, such as in healthcare or financial applications, where sensitive information must be protected while still enabling useful data analysis. The combination of PPkNN with SSED, SBD, and SMINn protocols provides a robust solution to the challenges of secure query processing over encrypted data in cloud environments.

## 2.1 Problem definition

Let's assume Sonia owns a database D of n records t1, ..., tn and m + 1 attributes. Let $t_{i,j}$ denotes the jth attribute value of record ti. Initially, Sonia encrypts her database attribute-wise, that is, she computes Epk(ti,j), for $1 \leq i \leq n$ and $1 \leq j \leq m+1$, where column (m+1) contains class labels. Assuming that the underlying encryption scheme is semantically secure. Let the encrypted database be denoted by D'. Also assuming that Sonia outsources D' as well as future classification processes to the cloud. Let Rajiv be an authorized user who wants to classify his input record q = hq1, ..., qmi by applying the k-NN classification method based on D'. Referring to such a process as privacy-preserving k-NN (PPkNN) classification over encrypted data in the cloud. We define the PPkNN protocol as: PPkNN(D', q) → cq, where, cq denotes class label for q after applying k-NN classification method on D' and q.

The program flow for this project is as follows:
Assume that p1 and p2 are two semi-honest parties such that the Paillier's secret key sk is known only to P2 whereas pk is public.

*Secure Multiplication (SM) Protocol*

This protocol considers P1 with input (Epk(a), Epk(b)) and outputs Epk(a*b) toP1, where a and b are not known to P1 and P2. During this process, no information regarding a and b is revealed to P1 and P2.

*Algorithm 1: SM(Epk(a), Epk(b)) → Epk(a * b)*
1: P1:

(a). Pick two random numbers ra, rb ∈ ZN.

(b). a' ← Epk(a) * Epk(ra)

(c). b' ← Epk(b) * Epk(rb)

(d). Send a', b' to P2

2: P2:

(a). Receive a', b' from P1

(b). ha ← Dsk(a'); hb ← Dsk(b')

(c). h ← ha * hb mod N

(d). h' ← Epk(h)

(e). Send h' to P1

3: P1:

(a). Receive h' from P2

(b). s ← h' * Epk(a)

(c). s ← s * Epk(b) ^ (N-rb)

(d). Epk(a * b) ← s * Epk(N - ra * rb)

*Secure Squared Euclidean Distance (SSED) Protocol:*
In this protocol, P1 with input (Epk(X),Epk(Y)) and P2 with securely kysk securely compute the encryption of squared euclidean distance between vectors X and Y. Here X and Y are m dimensional vectors where Epk(X)=(Epk(x1), Epk(x2)........., Epk(xm)) and Epk(Y)=(Epk(y1)…..,Epk(y2),.........., Epk(ym)). The output Epk(|X- Y|^2) will be known only to P1.

*Algorithm 2:* SSED(Epk (X), Epk (Y)) → Epk (|X- Y|^2)
Require: P1 has Epk(X) and Epk (Y); P2 sk

1: P1 for $1 \leq j \leq m$ do:

(a). Epk (xj – yj) ← Epk (xj) * Epk (yj) ^ N-1

2: P1 and P2 for $1 \leq j \leq m$ do:

(a). Compute Epk ((xj – yj)^2) using the SM protocol

3: P1:

(a). Epk ((X – Y)^2) ← ∏ Epk ((xj – yj)^2)

*Secure Bit-Decomposition (SBD) Protocol*
Here P1 with input Epk(z) and P2 securely compute the encryptions of the individual bits of z, where 0 <= z < 2^l. The output [z] = Epk(z1), ..., Epk(zl) is known only to P1. Here zl and z1 are the most and least significant bits of integer z, respectively.

*Algorithm 3:* Binary(x) → <x0,...,xm-1>
Require: A positive decimal integer x, where 0<=x<2^m

1. i←0

2. While i≠m do

3. xi←x mod 2

4. x←⌊x/2⌋ {observe that x is updated to current quotient qi}

5. i←i+1

6. end while

**3. Conclusion**

Referring to privacy-preserving k-Nearest Neighbor (PPkNN) classification over encrypted data in the cloud, we define the PPkNN protocol as: PPkNN(D', q) → cq, where cq denotes the class label for q after applying the k-NN classification method on D' and q. This research proposes a novel privacy-preserving k-NN classification protocol over encrypted data in the cloud. This protocol ensures the confidentiality of user data and query inputs while performing effective k-NN classification. The system employs advanced cryptographic techniques to maintain data privacy and prevent unauthorized access to sensitive information. The secure protocols incorporated—such as Secure Multiplication (SM), Secure Squared Euclidean Distance (SSED), Secure Bit-Decomposition (SBD), and Secure Minimum (SMIN)—work collectively on user's data/queries to process encrypted data securely. By outsourcing encrypted data and query processing to the cloud, users can benefit from reduced data management costs and enhanced computational efficiency. This approach is significant for applications where data privacy is critical, such as healthcare, finance, etc. The PPkNN protocol provides a robust solution to the challenges posed by traditional classification systems in untrusted cloud environments. It effectively mitigates the risks associated with data breaches and unauthorized access by ensuring that both the data and queries remain encrypted throughout the classification process. This research not only addresses the privacy concerns but also demonstrates the feasibility of performing complex data mining operations on encrypted data without compromising accuracy or efficiency.

In conclusion, the proposed PPkNN protocol represents a significant advancement in the field of privacy-preserving data mining, offering a practical and secure method for k-NN classification in cloud computing environments. Future work may explore optimizing these protocols for various cloud architectures and expanding their applicability to other machine learning algorithms.